\documentclass[aps,apl,twocolumn,showpacs]{revtex4-1}
\usepackage{graphicx}
\usepackage{amsmath}
\usepackage{amsfonts}
\usepackage{latexsym}
\usepackage{natbib}
\usepackage{bm}
\newcommand{\field}[1]{\mathbb{#1}}
\newcommand{\cs}{$\clubsuit$}

\newcommand{\I}{{\rm i}}
\newcommand{\e}{{\rm e}}
\begin{document}
\title{Equilibration and macroscopic quantum fluctuations in the Dicke model}

\author{Alexander Altland}

\affiliation{Institut f{\"u}r Theoretische Physik,
Universit{\"a}t zu K{\"o}ln, K{\"o}ln, 50937, Germany}

\author{Fritz Haake}

\affiliation{Fakult\"at f\"ur Physik, Universit\"at Duisburg-Essen,
  47048 Duisburg, Germany}

\begin{abstract}
  We discuss the unitary quantum dynamics of the Dicke model (spin and
  oscillator coupled). A suitable quasiprobabilty representing the
  quantum state turns out to obey a Fokker-Planck equation, with drift
  terms representing the underlying classical Hamiltonian flow and
  diffusion terms describing quantum fluctuations.  We show (by
  projecting the dynamics onto a co-moving Poincar\'e section) how the
  interplay of deterministic drift and quantum diffusion generates
  equilibration to the microcanonical density, under conditions of
  global classical chaos. The pertinent photon statistics reveals
  macroscopic quantum fluctuations.
\end{abstract}

\maketitle
\section{Introduction}

The Dicke model~\cite{Dicke:1954kx} has long been known for its quantum phase
transition (to a superradiant phase as the coupling exceeds a critical
value) \cite{Hepp:1973uq,*Hepp:1973fk} and classical
as well as quantum chaos~\cite{Graham:1984vn,*Graham:1984ys,
  *Graham:1986zr,Kus:1985ve,Emary:2003qf}. New interest in the model
has recently been stimulated by an experimental realization (employing
a double Bose-Einstein condensate coupled to an electromagnetic mode
of a surrounding cavity) where the zero-temperature phase transition was
observed~\cite{Baumann:2010vn}. Fascinatingly, a many-body system here
exhibits macroscopic quantum behavior involving only two coupled
degrees of freedom, one each for spin and oscillator.

The observation of chaos requires a combination of not too small
coupling and sufficient excitation. The experimental realization of
such regimes would reveal two distinctive signatures of chaos,
dissipationless equilibration on the energy shell~\cite{Altland:fk} and
macroscopic stationary fluctuations of the photon number, and both of
these should be accessible in the condensate/cavity setting of
Ref.~\cite{Baumann:2010vn}. For  large values of the
quantum number $j$ fixing the conserved square of the spin as
$\vec{J}^2=j(j+1)\gg 1$, the quantum dynamics will be
strongly influenced by the classical Hamiltonian flow. The essential
physics of the Dicke model will thus be governed by an interplay of
classical chaos and semiclassical quantum fluctuations.

The purpose of this paper is threefold:
\begin{itemize}
\item[i)] We will present a comprehensive discussion of chaos in the
  system. Unlike with the pioneering study \cite{Emary:2003qf},
 we will  not limit ourselves to the vicinity of the system's ground states but
  address the full phase space available to chaotic fluctuations.
\item[ii)] We discuss how a conspiracy of
  quantum fluctuations and nonlinear classical dynamics generates
  'thermalization' to a microcanonical distribution. 
  We thus present a first case study where the general role
  of quantum fluctuations in a chaotic thermalization process is
  resolved in microscopic detail. 
\item[iii)] We will analyse giant fluctuations of observables such as
  the photon number resulting from the uniform coverage of the
  system's energy shell.
\end{itemize}

\noindent i) For the classical motion of the Dicke model, we will
identify regimes of integrable, mixed and fully chaotic dynamics in
regimes of small, intermediate and strong spin-oscillator
coupling. Specifically, we will show that chaotic trajectories sweep
out large intervals of the classical action variable corresponding to
the number of oscillator quanta.  The tendency to uniform coverage of
the energy shell, typical for dominantly chaotic behavior, arises
already for rather moderate energies, provided the coupling is
sufficiently large. We conclude the classical discussion by
demonstrating that the classical Liouville equation would effectively
entail equilibration to the energy shell. The disclaimer 'effectively'
refers to the fact that the classical $Q$ evolves into an infinitely
rugged distribution of alternating high and zero phase space
density. Only in the infinite time limit, the ensuing structure
becomes infinitely filigree, and looks homogeneous at any finite
resolution.

\noindent ii) To decribe the quantum dynamics, we represent the
density operator by a suitable quasi-probability, the Glauber
$Q$-function aka Husimi function. Schr\"odinger's equation then turns
into a Fokker-Planck equation for $Q$. The drift terms therein
(first-order derivatives) reflect the classical Hamiltonian flow while
the second-order derivative terms, small in the parameter $1/j$,
describe quantum diffusion. The important role of quantum diffusion
is, under conditions of global classical chaos, to limit the shrinking
of phase space structures along the classically stable directions to a
scale of the order $j^{-1/2}$. The competition of classical vs quantum
contraction and expansion of phase space structures corroborates
equilibration to the microcanonical distribution.  In particular it
implies that a) any quasi-probability $Q$ will eventually cover the
compact energy shell of the model, on a time scale bounded from above
by the Ehrenfest time. Further, our theory reveals explicitly how
quantum diffusion smoothens the ruggedness of classical structures
over scales $\sim j^{-1/2}$. As a result, we obtain thermalization
into a genuinely uniform distribution (to be distinguished from a
fissured structure that only looks like a distribution upon decreasing
the level of resolution) on time scales of the order of the Ehrenfest
time $\sim \ln(j)$. It is worth emphasizing that the smoothing
mechanism relies on the effect of quantum fluctuations \textit{during}
the dynamical evolution, rather than just reflecting the
\textit{initial} quantum uncertainty intrinsic to any starting
distribution.

\noindent iii) The ensuing uniform coverage of the energy shell
entails large fluctuations of observables such as the photon number or
spin projections. The underlying (semiclassical) physics is that
during their dynamical evolution, system trajectories fully explore
the available phase volume in an essentially unpredictable manner. For
instance, in the superradiant regime and at energies $E \gtrsim \hbar
\omega_0 j$, where $\omega_0$ is the spin precession frequency, the
average photon number is roughly given by $\bar n \sim
\mathcal{O}(E/\hbar \omega)$. The fluctuations, $\delta n$
superimposed on this mean are of order $\delta
n\sim\mathcal{O}(j^{1/2}\sqrt{\bar n})\sim \mathcal{O}(\bar n)$,
i.e. 'macroscopic' fluctuations as big as the average may occur.

The rest of the paper is organized as follows. In
Sect.~\ref{sec:model} we discuss the Hamiltonian and the quantum
dynamics it generates. Specifically, the quasiprobability $Q$ will be
shown to obey a Fokker-Planck equation.
Sect.~\ref{sec:classical-dynamics} is devoted to the classical
dynamics and in Sect.~\ref{diffusion} we investigate quantum
diffusion. Finally, in Sect.~\ref{sec:macfluct} we discuss fluctuations of the photon number by calculating the
microcanonical averages of moments. In
Sect.~\ref{sec:beyond} we point out why the chaotic drift/diffusion
mechanism bringing about equilibration and large stationary
fluctuations works in other systems as well. Prominent examples are
the kicked top (recently realized experimentally) and the Bose-Hubbard
model (a genuine many-body system of much current interest).
Appendices will detail some calculations.

\section{Model and quantum evolution}
\label{sec:model}
The Hamiltonian of the Dicke model can be written as
\begin{equation}
  \label{H}
  \hat{H}=\hbar\Big\{\omega_0 \hat J_z+\omega a^\dagger a +
               g\textstyle{\sqrt{\frac{2}{j}}}(a+a^\dagger)\hat J_x\Big\}\,.
\end{equation} 
Here, the operators $\hat J_a, a=x,y,z$ act in a spin-$j$
representation and obey the standard commutation relations $[\hat
J_a,\hat J_b]=\I\epsilon_{abc} \hat J_c$, where $\{\epsilon_{abc}\}$ is
the fully antisymmetric tensor.  The photon annihilation and creation
operators fulfill the Bose commutation rules $[a,a^\dagger]=1$.  The
first two terms in \eqref{H} respectively describe spin precession
about the $J_z$-axis with frequency $\omega_0$ and harmonic
oscillation with frequency $\omega$. The last term accounts for spin
precession about the $J_x$-axis with a 'frequency' $\propto
a+a^\dagger$ and for driving of the oscillator by a 'force' $\propto
J_x$.  The coupling constant $g$ is a (Rabi) frequency independent of
$\hbar$. The appearance of the spin quantum number $j$ in the
interaction part is owed to the use of the operators $a,
a^\dagger,\vec{J}$ which are rather non-classical in character
\footnote{In particular, the photon operators $a,a^\dagger$ have no
  classical limit. If the Hamiltonian is written as
  $H=\frac{p^2}{2m}+\frac{1}{2}m\omega^2x^2 +\omega_0 L_z+\mu xL_x$
  all parameters and observables have well defined classical meanings;
  upon introducing $\vec{J}=\vec{L}/\hbar$ and
  $a=x\sqrt{\frac{m\omega}{2\hbar}}+i p\sqrt{\frac{1}{2\hbar
      m\omega}}$ we get the form (\ref{H}) with $
  g=\mu\sqrt{\frac{\hbar j}{4\omega m}}$.}.  We note that the
Hamiltonian (\ref{H}) contains the antiresonant terms $J_+a^\dagger
+J_-a$.  Only the parity $P=\exp \I\pi (a^\dagger a+J_z)$ thus remains
as a symmetry. If the antiresonant terms were dropped ('rotating wave
approximation'), conservation of $a^\dagger a+J_z$ and thus
integrability would result.

\subsection{Coherent state representation}
\label{sec:coher-state-repr}

We aim to explore the quantum dynamics generated by the Hamiltonian
\eqref{H}. In view of the largeness of the spin, $j\gg1$, we find it
convenient to employ coherent states which are optimally suited to
taking semiclassical limits.  Specifically, spin coherent states
\cite{Arecchi:1972kx,Glauber:1976uq,Scully:1994ly} are defined as
\begin{align}
  \label{eq:1}
  |z\rangle \equiv {1\over (1+|z|^2)^j} \e^{ z\hat J_-}\, |j,j\rangle,
\end{align}
where $|j,j\rangle$ is a 'maximum-weight' eigenstate of $J_z$, i.e.
$J_z |j,j\rangle=j |j,j\rangle$. The states $|z\rangle$ yield
the mean values
\begin{align}
  \label{eq:3}
  \langle z|\hat J_x|z\rangle &=j {z+z^*\over
    1+|z|^2}\equiv j l_x,\cr
\langle z|\hat J_y|z\rangle &={j\over \I} {z-z^*\over 1+|z|^2}\equiv j l_y,\cr
\langle z|\hat J_z|z\rangle &=j {1-|z|^2\over 1+|z|^2}\equiv j l_z;
\end{align}
here $l_{x,y,z}$ are the three components of a unit vector
$\mathbf{l}=(\sin\theta \cos\phi,\sin\theta\sin\phi,\cos\theta)^T$,
whose angular orientation is defined through $z=\e^{i\phi}
\tan(\theta/2)$. Higher moments reveal minimum angular uncertainty,
characterized by the solid angle $4\pi/(2j+1)$ which defines a Planck
cell on the unit sphere.  The set $\{|z\rangle \}$ is overcomplete and
allows the resolution of unity by $\openone=\int
dz\frac{2j+1}{\pi(1+zz^*)^2}|z\rangle\langle z|$.  Two spin coherent
states have the overlap $\langle z|z'\rangle=
\frac{(1+z^*z')^{2j}}{(1+|z|^2)^j(1+|z'|^2)^j}$.

Similarly,  oscillator coherent states are defined as \cite{Glauber:2007fk}
\begin{equation}
  \label{cohstateosc}
  |\alpha\rangle=\e^{-\alpha\alpha^*/2}\,\e^{\alpha a^\dagger}|0\rangle
\end{equation}
with $\alpha\in\field{C}$ a complex amplitude and $|0\rangle$ the
vacuum, $a|0\rangle=0$. The resolution of unity in terms of the
(over)complete set $\{|\alpha\rangle\}$ reads $\openone = {1\over
  \pi}\int d\alpha\, |\alpha \rangle \langle \alpha|$.  The state
$|\alpha\rangle$ assigns a minimal uncertainty product to displacement
and momentum such that these quantities are 'confined' to a single
Planck cell.  The latter property is also evident from the overlap
$|\langle\alpha|\alpha'\rangle|^2= \e^{-|\alpha-\alpha'|^2}$.

We shall discuss the dynamics of the system in terms of its density
operator $\hat \rho(t)$ and represent the latter by the Glauber $Q$ or
Husimi function \cite{Scully:1994ly}
\begin{align}
  \label{Q}
  Q(\alpha,z)=\frac{2j+1}{\pi (1+zz^*)^2}\langle \alpha,z|\hat\rho|\alpha,z\rangle\,.
\end{align}
Obviously, that function is real and non-negative everywhere and
exists for any density operator $\hat\rho$.  By invoking the completeness
relations given above, one immeadiately checks that expectation values
of (anti-normal ordered) operators of the oscillator can be computed
as
\begin{equation}
  \label{moments}
  \langle a^m a^{\dagger n}\rangle=
                \int d\alpha dz\, \alpha^m{\alpha^*}^{n}\,Q(\alpha,z)\,.
\end{equation}
%In passing we note that 
Expectation values of spin operators 
can be computed by analogous averaging over
$z$-valued functions. We just note the example 
\begin{equation}
\label{mixedmoments}
 \langle a^m a^{\dagger n}\hat J_a\rangle=
            \int d\alpha dz\, \alpha^m{\alpha^*}^{n}
              (j+1)l_a(z,z^*)%+\mathcal
           %O(\textstyle{\frac{1}{j}})\Big) 
           Q(\alpha,z)
\,, 
\end{equation}
where the unit increment over $j$ in the factor $(j+1)$ is a quantum
correction. At any rate, the foregoing properties of $Q$ allow us to
speak of a quasi-probability density which is expected to converge to a classical
phase-space density in the limit $\hbar \to 0$.

To illustrate the use of our quasiprobability we note that a density
operator projecting onto a pure coherent state $|z_0,\alpha_0\rangle
\equiv |z_0\rangle \otimes |\alpha_0\rangle$ implies a $Q$-function
smeared out over a single Planck cell as
\begin{eqnarray}
  \label{initialcohstate}
Q(z,\alpha) &=&\frac{2j+1}{\pi^2(1+|z|^2)^2}\,\big|\langle
\alpha,z|\alpha_0,z_0\rangle\big|^2\,.%\\
%\nonumber
%&=&\frac{1}{\pi}e^{-|\alpha-\alpha'|^2}\frac{2j+1}{\pi(1+|z|^2)^2}
%       \left|\frac{(1+z^*z')^{2}}{(1+|z|^2)(1+|z'|^2)}\right|^{2j}\,.
\end{eqnarray}
We shall in fact mostly imagine the system initially prepared in such
a nearly classical state.

\subsection{Time evolution}
\label{time evolution}

The von Neumann equation $d_t\hat \rho=\frac{\I}{\hbar}[\hat
H,\hat\rho]$ can be rewritten as a partial differential equation for
the quasi-probability $Q$. Using the definition of the coherent states
we find (see appendix \ref{sec:deriv-quant-evol}) that equation to
involve only first and second derivative terms respectively
interpretable as classical drift and quantum diffusion,
\begin{widetext}
\begin{align}
  \label{quantumevol}
  \dot{Q}&=(\mathcal{L}+\mathcal{L}_{\rm diff})Q\,\\ \nonumber
  \mathcal{L} &=\I\partial_\alpha\left(\omega\alpha + 
                g\sqrt{\frac 2 j}(j+1)\frac{z+z^*}{1+|z|^2} \right)
          +\I\partial_z\left(\!\!-\omega_0 z+
                 \frac{g}{\sqrt{2j}}(1-z^2)(\alpha+\alpha^*)\!\right)
               +{\rm c.c.}\,,
            \quad   
   \mathcal{L}_{\rm diff} = \frac{\I g}{\sqrt{2j}}\partial_{\alpha}\partial_z  (1-z^2)
                     +{\rm c.c.}\,,
\end{align}
\end{widetext}
and may thus speak of a Fokker-Planck equation. We add in passing that
we disregard any damping, restricting ourselves, with respect to the
condensate/cavity experiment of Ref.~\cite{Baumann:2010vn}, to times
smaller than the life times of both the cavity photons and the
condensate.

To discuss the semiclassical limit\footnote{We set $(j+1)\to j$ in the
  oscillator drift since the unit increment, of the same origin as the
  one in (\ref{mixedmoments}), has no further interest in the
  semiclassical limit. That increment would be relevant only for
  next-to-leading order corrections to means (of products) of the $J_a$.}
$j\gg1$, it is convenient to switch to variables obeying canonical
classical commutation relations. For the oscillator, we introduce 'action-angle'
variables $(I,\psi)$ through \footnote{Notice that our dimensionless
  action variable $I$ differs from the standard oscillator action
  variable, $\tilde I=a^\dagger a \hbar$, by a constant factor, $I =
  \tilde I j/\hbar = \tilde I L$, where $L$ is the conserved value of
  the classical spin angular momentum.}
\begin{align}
  \label{eq:8}
\alpha=\sqrt{jI} \e^{\I\psi},\quad \alpha^* = \sqrt{jI}
\e^{-\I\psi}.  
\end{align}
To parametrize the Bloch sphere of the spin we employ
$(\cos\theta,\phi)$, cf. \eqref{eq:3}.  Both pairs are canonical, with
Poisson brackets $\{I,\psi\}=\{\cos\theta,\phi\}=1$. A few details
pertaining to the change of variables are given in Appendix
\ref{sec:trafo}.

Expressed in terms of these variables, the drift operator
$\mathcal{L}$ assumes the form of a Liouvillian,
$\mathcal{L}=-\{h,\;\}$, 
with the effective Hamiltonian function
\begin{align}
  \label{eq:7}
  h=\omega_0 \cos\theta + \omega I + g \sqrt{8I}
    \cos\psi \sin\theta \cos\phi,
\end{align}
obtained from the Hamilton operator \eqref{H} by substituting 
$\hat{\mathbf{J}}\to j \mathbf{l}$, $a\to \sqrt{jI} \exp(\I\psi)$,
and dividing out $\hbar j =L$. The classical approximation
$d_t Q \simeq \mathcal{L}Q=-\{h,Q\}$ to the evolution equation
(\ref{quantumevol}) describes a drift of the quasi-probability
$Q(I, \psi, \cos\theta, \phi)$ along the classical trajectories of the
Hamiltonian flow. The latter are determined by the Hamiltonian
equations of motion
\begin{align}
  \label{Hameqs}
  \dot{I}&=&-\sqrt{8}g\sqrt{I}\sin\psi\sin\theta\cos\phi\\
\nonumber
  \dot{\psi}&=&-\omega-\sqrt{2}g\frac{1}{\sqrt{I}}\cos\psi\sin\theta\cos\phi
\\ \nonumber
 \dot{\phi}&=&\omega_0-\sqrt{8}g\sqrt{I}\cos\psi\cot\theta\cos\phi
\\ \nonumber
  \dot{\cos\theta}&=&\sqrt{8}g\sqrt{I}\cos\psi \sin\theta\sin\phi\,.
\end{align}
As befits classical Hamiltonian equations the quantum number $j$
does not show up here.

When the quantum diffusion operator $\mathcal{L}_{\rm diff}$ is
written in terms of the above canonical pairs of variables, it
acquires a pre-factor $\frac{1}{j}$ which is very small in the
semiclassical limit. That semiclassical smallness notwithstanding,
quantum diffusion has an important smoothing effect on the
quasi-probabilty $Q$, as we shall see presently. But first, we devote
a thorough discussion to the drift.

\section{Classical dynamics}
\label{sec:classical-dynamics}

In this section, we will analyse the phase space flow according to the
classical Hamiltonian equations \eqref{Hameqs}, in regimes of
integrable, chaotic, and mixed dynamics.  A discussion of chaos in the
system has been reported in a seminal paper by Brandes and
Emary~\cite{Emary:2003qf}. However, the Holstein-Primakoff bosons
employed to represent the spin variables in that reference tend to
obscure the large-scale phase-space structure of the problem, and
notably the semiclassical limits $j\to \infty$ and
$\hbar \to 0$ at $\hbar j\equiv L=\mathrm{const}$.  For an insightful
discussion of chaos in the equations of motion
\eqref{Hameqs} (applied to an opto-mechanical setting) we also refer
to Ref.~\cite{Larson:2010uq}. The primary objective of our classical
analysis is to set the stage for the discussion of the quantum ramifications
of chaos.

\subsection{Qualitative picture}
\label{sec:qualitative-picture}

The equations of motion \eqref{Hameqs} do not involve the quantum
parameters $\hbar$ and $j=L/\hbar$. Remarkably, the scaled Hamiltonian
$h$ is also independent of the classical angular momentum $L=j\hbar$,
due to the particular scaling \eqref{eq:8} of the oscillator
variables. Put differently, the magnitude of the classical angular
momentum does not affect the dynamics and can be accommodated in a
rescaling of variables. The dynamics then depends on the dimensionless
measures for frequency, coupling, and energy
\begin{equation}
\nu\equiv {\omega\over \omega_0},\qquad \gamma \equiv{g\over g_c},\qquad \epsilon\equiv h/\omega_0\,,
\end{equation}
with $g_c\equiv \sqrt{\omega\omega_0}/2$ the critical coupling at
which the Dicke model undergoes its transition to a superradiant
phase. Without much loss of generality, we will assume comparable
frequencies $\omega/\omega_0=\mathcal{O}(1)$ throughout\footnote{To
  thoroughly invalidate the rotating wave approximation, the detuning
  $|\omega-\omega_0|$ must exceed the natural widths of the energy
  levels involved.}.
\begin{figure}
  \centering
  \includegraphics[width=6.5cm]{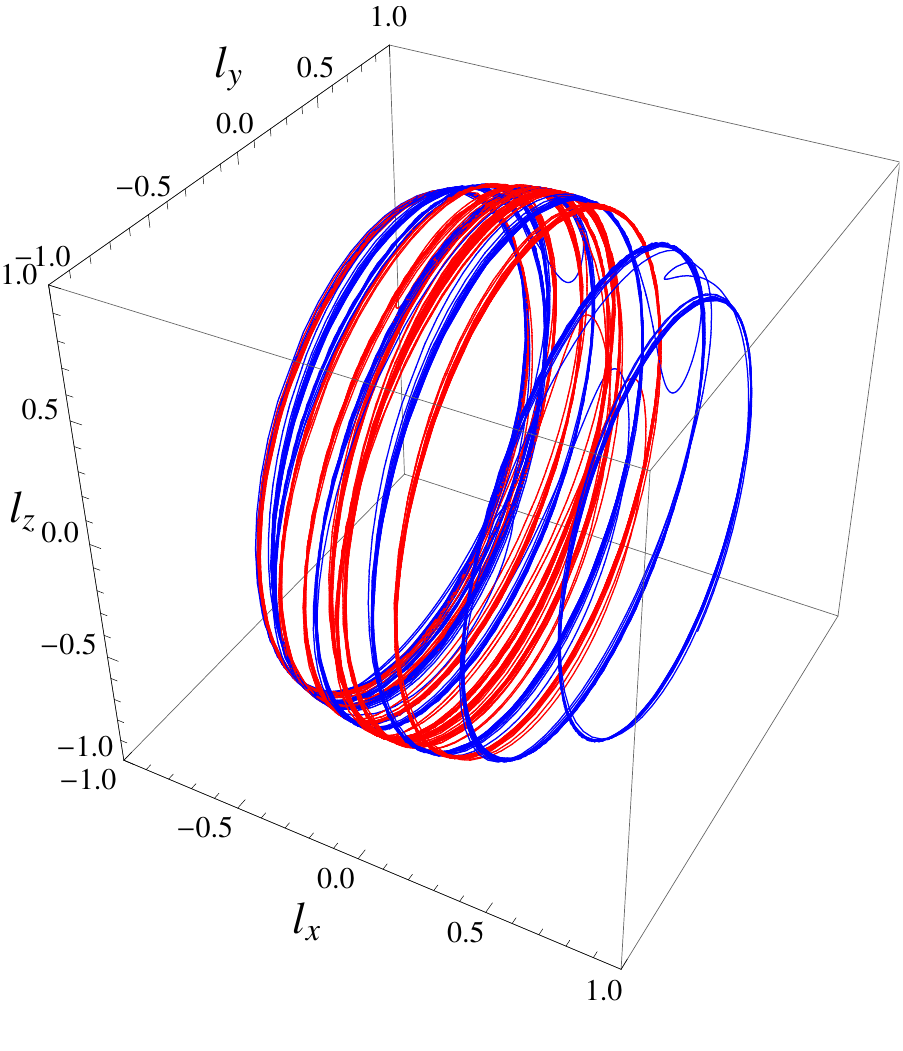}
  \includegraphics[width=8cm]{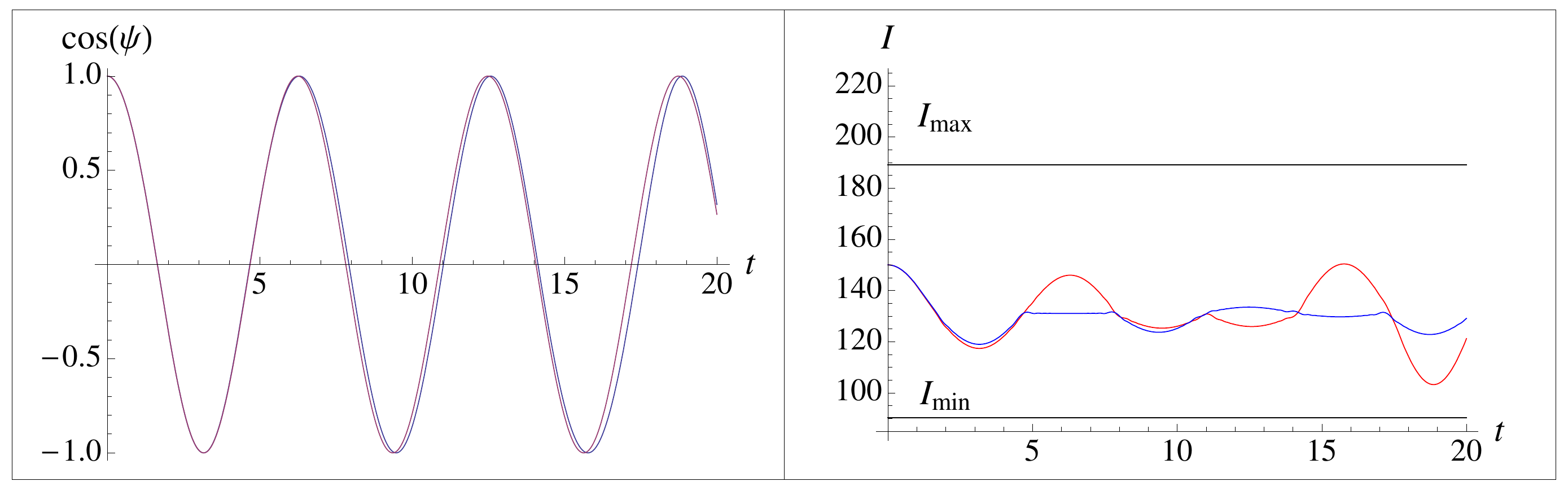}
  \caption{Chaotic trajectories of the classical Dicke model at $g=3
    g_c$  and $h\simeq 150 \omega_0$. The red and the blue trajectory
    differ in a slight mismatch of the initial conditions.}
  \label{fig:trajectory}
\end{figure}
To obtain some intuition of the dynamics, let us express the
Hamiltonian in terms of the unit-length angular momentum
$\mathbf{l}=(l_x,l_y,l_z)$:
\begin{align}
  \label{eq:10}
 {h\over \omega_0} = l_z+\nu I  + \gamma\sqrt{2\nu I}\cos\psi l_x\,,
\end{align}
and consider  high energy $\epsilon\gg 1$ and  strong
  coupling $\gamma>1$.  The time variation of the oscillator phase is
given by $\dot \psi = -\omega- g\sqrt{2/I}\cos \psi l_x = \omega +
\mathcal{O}(I^{-\frac{1}{2}})$. For sufficiently large $I$, and times
larger than the oscillation period $\sim \omega^{-1}$ of $\psi$, the
second term becomes negligible, i.e. we may approximate $\cos\psi
\simeq \cos(\omega t + \psi(0))$. The nearly harmonic oscillation of
$\cos\psi$ is exemplified by the trajectory shown in
Fig. \ref{fig:trajectory} (bottom left inset). It implies that the
model behaves, at the high energies under consideration, much like a
system of \textit{three} dynamical variables $(I,\theta,\phi)$ subject
to external harmonic driving $\sim \cos\psi$ at a frequency
$\omega$. The spin dynamics 
is governed by fast precession of the angular momentum $\mathbf{l}$
around the instantaneous rotation axis $\bm{\Omega}\equiv
\omega_0(\gamma\sqrt{2\nu I} \cos\psi,0,1)^T$. For 'typical' values of
$\cos\psi$, the precession frequency $|\bm{\Omega}|=\omega_0\sqrt
{(\gamma^2 2\nu I (\cos\theta)^2+1)}=\mathcal{O}( \omega_0\sqrt{I})\gg
\omega$ exceeds the 'driving' frequency $\omega$ by far.  This is
visible in the fast spinning of the variable $\mathbf{l}$ around the
unit sphere shown in the top section of
Fig.~\ref{fig:trajectory}. Second, the precession axis is typically
oriented in $x$-direction, $\bm{\Omega} \simeq \omega_0\gamma
\sqrt{2\nu I} \cos\psi \mathbf{e}_x +\mathcal{O}(1/\sqrt{I})$, and
during these periods the angular momentum component $l_x$ is
approximately conserved. This latter fact has important consequences
for the variation of our primary variable of interest, $I$. Over time
intervals of nearly conserved $l_x$, the equation of motion $\dot I =-
\omega_0 \gamma\sqrt{2\nu I} \sin\psi l_x$ can be trivially integrated
to obtain the characteristic arcs visible in the bottom right panel of
Fig.~\ref{fig:trajectory}. For any particular energy set by the
initial condition, the action variable varies between an upper and a
lower bound (indicated by horizontal
lines) %which can be obtained by a
calculated %ion detailed
in Appendix \ref{bounds}.
%(see also the discussion below.)

At times $t\sim (n\pi + \pi/2)/\omega$ (the $\sim$ indicates a jitter
of the order $1/\sqrt{I}$) the regular pattern outlined above gets
interrupted, when the phase $\cos(\omega t)$ becomes small enough for
$\Omega_x\sim \cos (\omega t)$ and $\Omega_z$ to be comparable. During
these short time spans, the angular momentum precesses around a vector
$\bm{\Omega}$ no longer aligned in $x$-direction to a new orientation
(cf. the isolated arcs visible in the top part of
Fig.~\ref{fig:trajectory}.) Specifically, the $x$-component $l_x$
changes to a new and essentially un-predictable value. After the time
window of small $\cos\psi$ has been left, $l_x$ is approximately
conserved again and the near regular change of $I$ re-commences, at a
changed rate $\sim l_x$.

\begin{figure}
  \centering
  \includegraphics[width=6.5cm]{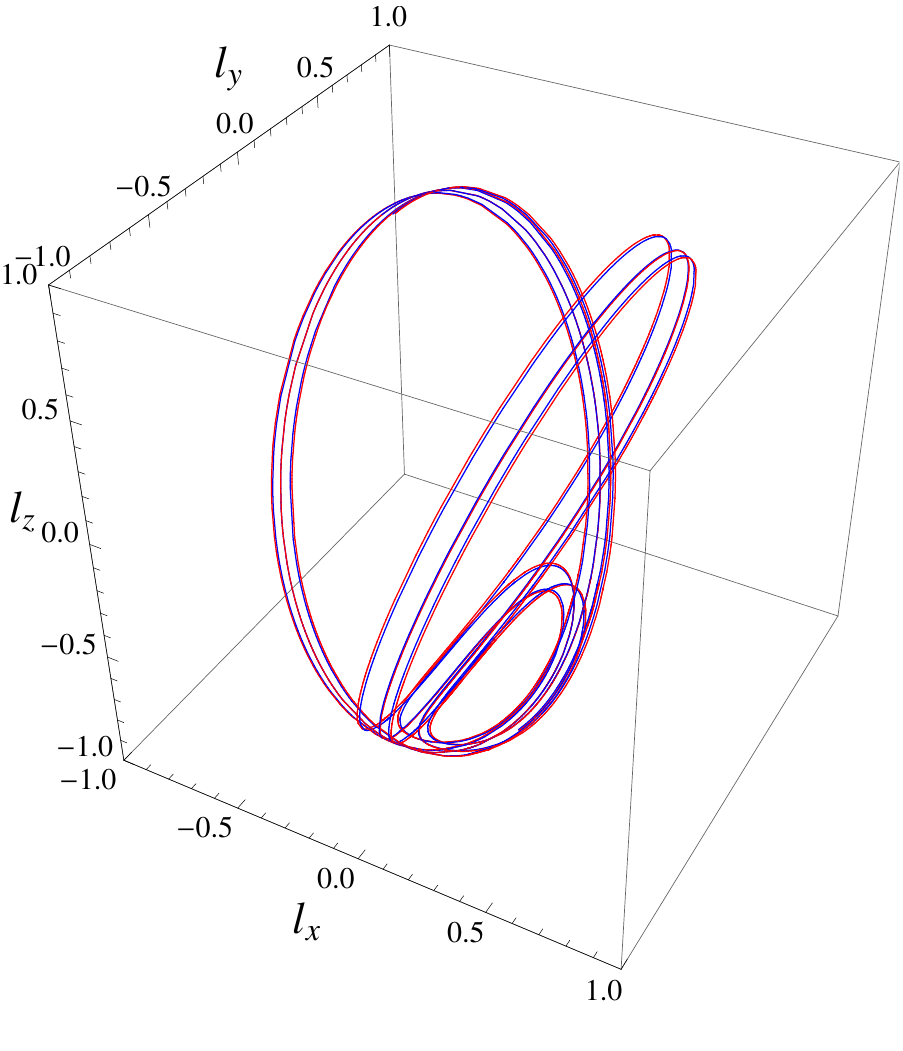}
  \includegraphics[width=8cm]{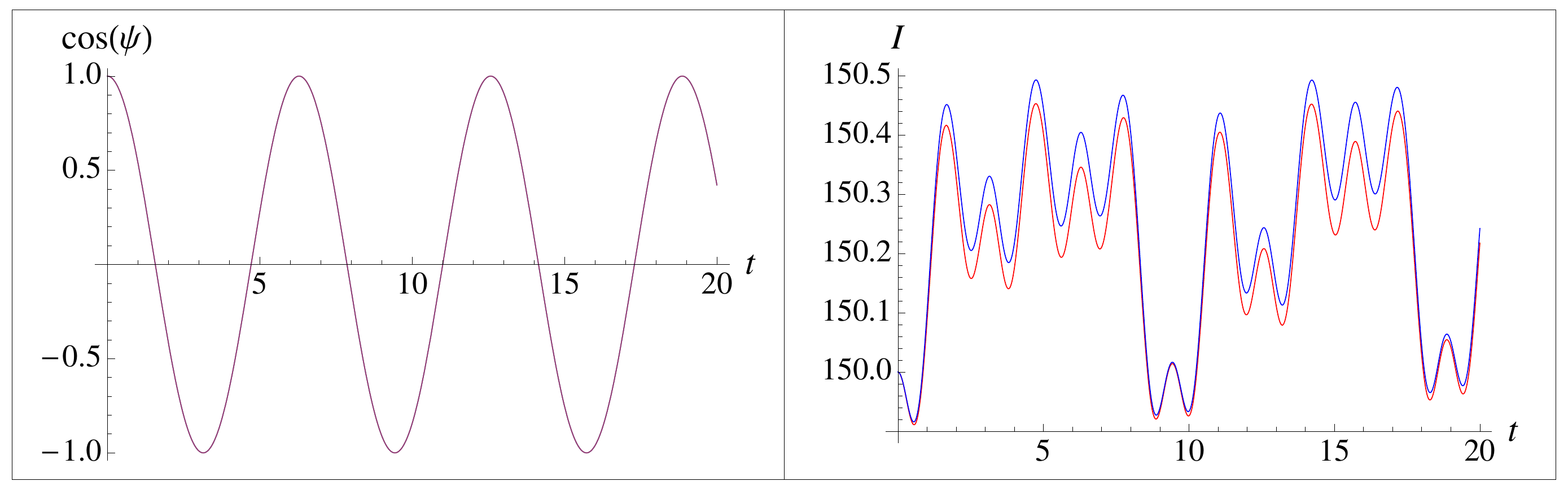}
  \caption{Chaotic trajectories of the classical Dicke model at $g=0.2
    g_c$ and $h\simeq 150 \omega_0$. The red and the blue trajectory
    differ in a slight mismatch of the initial conditions. \cs Note
    the scale on which $I$ varies, tiny compared to the chaotic case
    of Fig.~\ref{fig:trajectory}.}
  \label{fig:trajectoryI}
\end{figure}

Summarizing, the system behaves as if influenced by a 'random
number generator': at regular time steps $t\sim (n\pi +
\pi/2)/\omega$, a new value of $l_x$ is dialed up, and that value sets
the rate at which $I(t)$ changes during the consecutive time interval
of duration $\pi/2\omega$. Chaos manifests itself in this process's 
sensitivity to initial conditions. In the Fig.~\ref{fig:trajectory}
this is exemplified in terms of two trajectories of slightly different
initial value of the coordinate $I$. By
comparison, Fig.~\ref{fig:trajectoryI} visualizes the profile of
trajectories in the integrable realm. Notice the strongly
reduced, and effectively periodic fluctuations of the action variable,
and the lack of divergence of trajectories of different initial conditions. 

\subsection{Crossover to chaos}

Even though the literature on the Dicke model is vast no conclusive
treatment is available of the emergence of chaos as energy and
coupling strength are varied.  Filling that gap appears all the more
desirable as much previous work is focussed on low energies where the
flow cannot explore all of the spin sphere. The relative status of the
superradiant phase transition at $g=g_c$ and the crossover from
regular to chaotic behavior could thus not be reliably ascertained.

  \begin{figure*}
  \centering
  \includegraphics[width=18cm]{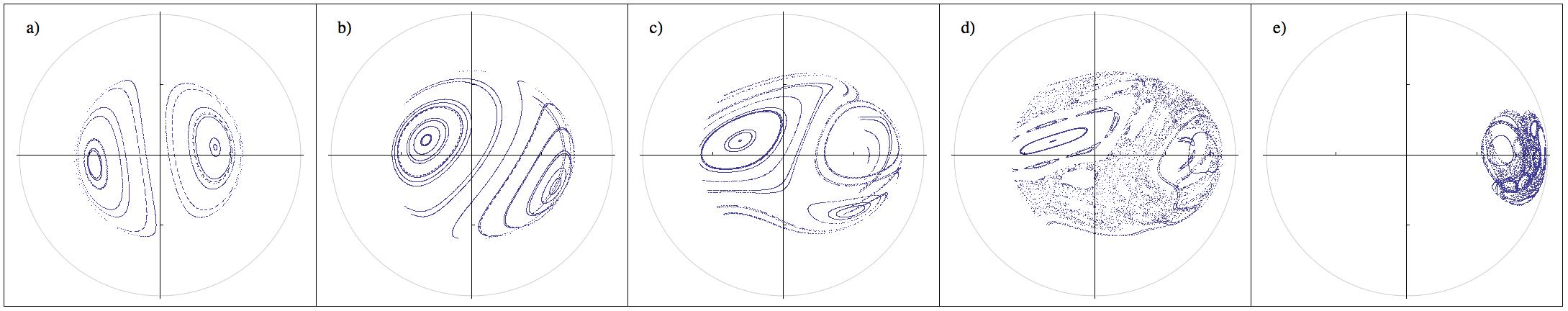}
   \caption{Poincar\'e sections generated by monitoring the pair
    $(l_x,l_y)$ at fixed values of the phase $\psi$, and at energies
    $\Delta \epsilon=0.2|\epsilon_0|$ above the ground state; only
    southern hemisphere shown since northern one remains empty. For each
    parameter value, $\gamma$, nine trajectories of different on-shell
  initial conditions are sampled. a) $\gamma=0.2$, b) $\gamma=0.7$,
  c) $\gamma=0.9$, d) $\gamma=1.01$, e) $\gamma=1.5$. What looks like
     crossings in (b, c) is a reflection of the non-uniqueness of
    $I(\epsilon,\psi,\cos\theta,\phi)$, see App. B. Noteworthy is the
    predominance of chaos in (e), for an energy as small as 20\% of
    the maximum energy capacity of the spin.}  
   \label{fig:poincarelow}
  \end{figure*}

Global chaos is prevalent at large energies $\epsilon\gg1$
and strong coupling $\gamma>1$. In general, the dynamics is mixed, or,
in limiting cases, integrable. To map out the regimes of different
dynamical behavior, we separately consider the model at weak and high
excitation energies.

\textit{Low energy dynamics .---} At weak coupling $\gamma<1$, the
Dicke Hamiltonian possesses a stationary point of lowest energy
$\epsilon_0=-1$ at $I=0$, $\cos\theta=-1$. In the energetic vicinity
of this point, the dynamics is integrable. Here 'vicinity' means
excitation energies $\Delta \epsilon\ll 1$, where
$\epsilon\omega_0=2\omega_0$ defines the maximum %(dimensionfull)
energy that can be accommodated by the spin. Integrability is visible
in the Poincar\'e sections shown in the first few panels of
Fig. \ref{fig:poincarelow}. Signatures of mixed dynamics become
visible upon approaching the critical value $\gamma=1$. At $\gamma>1$,
the ground state configuration shifts to an energy $\epsilon_0 =
-(\gamma^2+\gamma^{-2})/2$, which is now attained at two
degenerate points $(I,\psi)={1\over
  2\nu}(\gamma^2-\gamma^{-2}),0/\pi)$ and
$(\cos\theta,\phi)=(-\gamma^{-2},\pi/0)$; here the non-vanishing value
of the action coordinate corresponds to a macroscopic photon number
$\langle a^\dagger a\rangle = j I$. In the immediate vicinity of these
points, the dynamics remains integrable, and for moderate excitation
it is mixed. The subsequent crossover to chaotic dynamics
turns out to be rather swift; already at excitation energies $\Delta
\epsilon\simeq 0.2 |\epsilon_0|$, the energy shell is filled by
chaotic trajectories.

\begin{figure*}
  \centering
  \includegraphics[width=18cm]{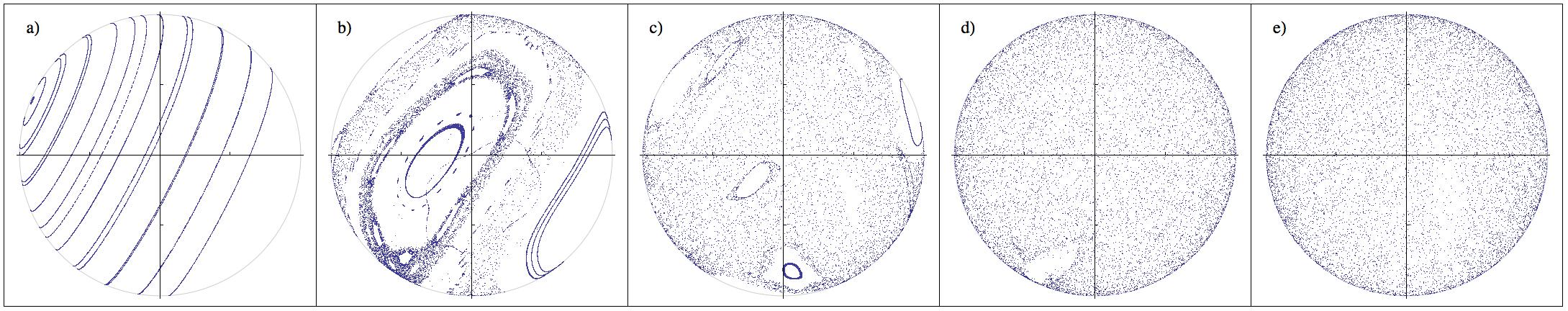}
  \caption{Poincar\'e sections as in Fig. \ref{fig:poincarelow}, now
    sampled at large energies $\Delta
    \epsilon=30|\epsilon_0|$. %Configurations $(l_x,l_y)$ on the
    %southern and northern ($l_z<0$ and $l_z>0$) hemisphere are shown
    %in the upper and lower panel. %a) $\gamma=0.2$, b) $\gamma=0.7$, c)
    %$\gamma=0.9$, d) $\gamma=0.97$, e) $\gamma=1.01$
    Values of $\gamma$ as in previous figure. The projection of the
    northern hemisphere, now fully covered by trajectories, looks qualitatively similar.}
   \label{fig:poincarehigh}
  \end{figure*}

\textit{High energy dynamics .---} At large energies, $\epsilon\gg 1$,
the Bloch sphere gets fully covered by trajectories. Already at
coupling strengths $\gamma<1$, trajectories become chaotic. The last
tori get lost in the immediate vicinity of the critical value $\gamma=1$,
cf. Fig. \ref{fig:poincarehigh}.

  For fully developped chaos, typical trajectories explore all of the
  energy shell $\epsilon\omega_0=h$. In view of our later discussion
  of photon number fluctuations, we need to explore the confines of
  these shells, especially with regard to the coordinate $I$. Lower
  and upper bounds of the action variable can be obtained as a result
  of a straightforward calculation detailed in Appendix
  \ref{bounds}. For large excitation energies, the accessible window
  of $I$-values asymptotes to
\begin{equation}
  \label{boundsasymp}
  I_{\rm max\atop\rm min}=\frac{\epsilon}{\nu}\left(1\pm\sqrt{\frac{2\gamma^2}{\epsilon}}\right),
\end{equation}
with corrections of $\mathcal{O}(1/\epsilon)$.  We are, thus, facing a
window of width $\propto \sqrt{\epsilon}$ and center $\propto
\epsilon$. In Section~\ref{sec:macfluct}, we will consider the
ramifications of the ergodic filling of these windows in the quantum
dynamics of the system.

\subsection{Evolution of $Q$ under chaotic drift}
Focusing on global chaos from now on, we would like to clarify how the
$Q$-function would evolve if quantum diffusion were neglected. The
assumed largeness of $j$ indeed suggests to try out that most radical
implementation of the semiclassical limit.

For an initial coherent state, $Q$ is isotropically 'supported' by a
single Planck cell, i.e., by a tiny fraction of the energy shell
located at $\langle\alpha_0,z_0|\hat
H|\alpha_0,z_0\rangle=\hbar\omega_0\epsilon\propto j$ with width
$\hbar\omega_0\Delta\epsilon=[\langle\alpha_0,z_0|(\hat H
-\hbar\omega_0\epsilon)^2|\alpha_0,z_0\rangle]^{1/2}\propto
\sqrt{j}$. By Liouville's theorem, that tiny fraction will not change
in time. The chaotic evolution will interminably squeeze the
originally 'circular support' of $Q$ in the stable direction of the
Hamiltonian flow and stretch it along the unstable direction. The ever
narrower and longer 'supporting stripe' must soon begin to fold since
the energy shell is compact. After a time of the order of the
Ehrenfest time the stripe will have fully explored the energy
shell. As the squeezing/stretching/folding of the support of $Q$
continues, an ever finer, and eventually 'singular' structure arises,
where $Q$ alternates infinitly rapidly between high and near vanishing
values transverse to the supporting stripe. Inasmuch as no region
within the energy shell appears favored, a constant mean density will
arise. If one were to look at the 'fissured landscape' formed by $Q$
with finite resolution one would, from a certain time on, just observe
'flatness' at the mean value of $Q$ mentioned. In other words, one
would see the microcanonical distribution: $Q$ constant within and
zero outside the energy shell. Expectation values of observables like
low-order powers of the photon number $\langle
(a^{\dagger}a)^m\rangle$ will not register the ruggedness of $Q$ but
just 'pick up' the microcanonical shape. Somewhat cavalierly said, $Q$
effectively equilibrates to the microcanonical distribution, within a
time of the order of the Ehrenfest time.

The foregoing scenario changes little if we imagine the initial
coherent state replaced by a squeezed minimum-uncertainty
state. Initial states with larger uncertainties bring but two changes:
(i) equilibration will happen even faster, the time scale shrinking
logarithmically with the initial width, and (ii) the landscape
underlying the effectively microcanonical $Q$ can be smoother.

\section{Quantum diffusion}
\label{diffusion}

Still confining ourselves to global chaos we now proceed to studying
how quantum diffusion changes the effective equlibration just found
for the classical drift. We shall find a smoothing effect of quantum
fluctuations which becomes effective, roughly, at phase space length
scales $\sim \sqrt{\hbar}$. To the best of our knowledge, our analysis
of the Dicke system represents the first case study where the
interplay of quantum fluctuations and nonlinear dynamics in the long
time behavior of a chaotic quantum system is resolved in concrete
terms.  A glance at the quantum diffusion operator (\ref{quantumevol})
reveals that $\mathcal{L}_{\rm diff}$ couples oscillator variables to
spin variables but does not include second-order derivatives wrt only
oscillator variables nor wrt only spin variables. That structure is of
course preserved when the canonical pairs $(I,\psi)$ and $(\cos\theta,
\phi)$ are employed, as we imagine done here. A real symmetric
$4\times 4$ diffusion matrix then arises which has vanishing
'diagonal' $2\times 2$ blocks and mutually Hermitian conjugate
'off-diagonal' $2\times 2$ blocks,
%\begin{equation}
 % \label{diffmatrix}
 $D=\left({0\atop d^\dagger}{d \atop 0}\right)$\,.
%\end{equation}
We shall not need the explicit dependence of the off-diagonal blocks
$d,d^\dagger$ on the variables $I,\psi\cos\theta, \phi$ here but would
like to emphasize the smallness $d\propto \frac{1}{j}$. 

The 'chiral' block structure of $D$ entails a secular equation for the
eigenvalues of the form $\lambda^4-\lambda^2\mbox{
  tr}\,dd^\dagger+\det{dd^\dagger}=0$. The four eigenvalues of $D$
thus come in two plus/minus pairs $\pm D_1,\pm D_2$ where
$D_1^2,\,D_2^2$ are the eigenvalues of the non-negative $2\times 2$ matrix
$dd^\dagger$. Each of these pairs is associated with an eigenvector
pair defining a contractive $(-)$ resp. expansive $(+)$ direction. In
the expansive directions we confront normal diffusion while for the
contracting directions we may speak of anti-diffusion. 

\subsection{Qualitative discussion of equilibration}
The quantum diffusive contraction (expansion) competes, given chaos,
with the stretching (shrinking) inherent in the classical
drift. Quantum antidiffusive shrinking will be overwhelmed by the
exponential deterministic expansion in the
classically unstable direction: The pertinent scales will keep
growing and the corresponding structure will ever more ubiquitously
explore the energy shell, much as if quantum diffusion were entirely
absent. However, in the deterministically stable direction where
exponential shrinking proceeds ever more slowly, quantum
diffusion will not allow that shrinking to go below a quantum scale $\propto
\frac{1}{\sqrt{j}}$.  Therefore, the $Q$-function will be diffusively
smoothed transverse to the unstable direction such that the
deterministically favored fissured 'landscape' never
arises. Microcanonical flatness will be reached from any initial
state, coherent, squeezed, or broader, on the Ehrenfest time scale or
faster.

The foregoing arguments do not rule out revival events, but such
cannot be expected any earlier than a Heisenberg time $t_H\propto j$, 
possibly even a Poincar\'e time ($\propto \e^j$).

\subsection{Quantitative discussion: Co-moving quantum fluctuations} 

The picture just drawn faithfully reflects a systematic theory
obtained by projecting the dynamics onto a co-moving Poincar\'e
section. To see that we pick a phase space point $X_0$ and a
deflection $\delta X$. The latter may be expressed through
increments of the canonical variables, $\delta X=(\delta I, \delta
\psi, \delta \cos\theta, \delta \phi)^T$, or through components
$(s,u,\epsilon,\tau)^T\equiv \xi$ along the four directions
distinguished asymptotically by the classical flow: one stable, one
unstable, and two neutrals (transverse to the energy shell and along
the flow). The two variants are linearly related,
\begin{equation}
  \label{increments}
  (\delta p, \delta x, \delta\cos\theta, \delta\phi)^T=C (s,u,\epsilon,\tau)^T\,,
\end{equation}
or in brief $\delta X=C\xi$.  The $4\times4$ matrix $C$ is composed by
the unit vectors $e_s,e_u,e_\epsilon,e_\tau$ along the
stable/unstable/neutral directions as $C=(e_s,e_u,e_\epsilon,e_\tau)$;
it depends on $X_0$ and must be determined numerically.

\begin{widetext}

  \begin{figure}[h]
  \centering
  \includegraphics[width=15cm]{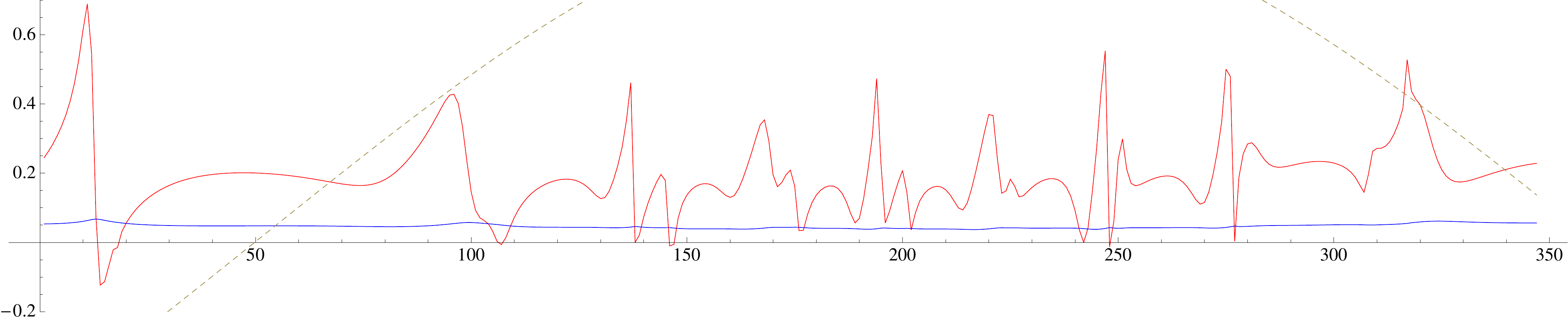}
  \caption{ \label{fig:Overlap} (Color online) Red: the coefficient $D_{ss}$, Eq.~\eqref{dss}, numerically
    computed along a trajectory of energy $h\simeq 250 \omega_0$ and
    $\alpha=1.1$. Blue: $D_{ss}$ integrated against the Lyapunov
    kernel $\exp(-2\lambda t)$ as in \eqref{stablevariance}. In the
    (arbitrary) units of the Plot, $\lambda^{-1}\simeq 150$. Dashed: the 'clock-variable'
  $\cos(\psi)$ effectively measuring time in units of
  $\omega^{-1}$.} 
\end{figure}
\end{widetext}

Now we turn to the Fokker-Planck equation for $Q$. %$\dot
%Q=\big(\partial_{X_i}d_i(X)+\partial_{X_i}\partial_{X_j}D_{ij}(X)\big)Q$
Near $X_0$ the components of $\delta
X$ can be employed as $\dot Q=\big(\partial_{\delta
  X_i}d_i(X_0+\delta X)+\partial_{\delta X_i}\partial_{\delta
  X_j}D_{ij}(X_0+\delta X)\big)Q$ with $d(X)$ the drift 'vector' and
$D(X)$ the diffusion matrix.  We then  change variables according to
(\ref{increments}) and $\partial_{\delta X}=C^T\partial_\xi$. The
drift can be linearized and has the first two components $d_s=\lambda
s,\; d_u=-\lambda u$ with $\lambda$ the Lyapounov rate. On the other
hand, in the diffusion matrix we may drop the deflection $\xi$. Then
integrating out the variables $\epsilon,\,\tau$ we get the bivariate
density $Q(s,u)$ which obeys the reduced Fokker-Planck equation
\begin{equation}
  \label{reducedFP}
  \dot Q=\big(\lambda\partial_s s-\lambda\partial_u u+
               \partial_s^2D_{ss} +\partial_u^2D_{uu}+2 \partial_s\partial_uD_{su}\big)Q\,.
\end{equation}
Here the 'reduced diffusion matrix' matrix $\left({D_{ss}\atop
    D_{us}}{D_{su}\atop D_{uu}}\right)$, describing quantum diffusion
in the stable/unstable subspace, is the upper left $2\times2$ block in
$ C^TDC$. Intuitive expressions arise when the orthonormal
eigenvectors $v_\mu$ and eigenvalues $D_\mu$ of $D$ are used, like
\begin{equation}
  \label{dss}
D_{ss}=\sum_\mu(\langle e_s, v_\mu\rangle)^2D_\mu\,.
\end{equation}
The sign of that effective diffusion constant will be of upmost
importance. We see that the relative orientation of the intervening
vectors matters, as well as the presence of positive and negative
eigenvalues $D_\mu$.

Next, we proceed from the local quantum fluctuations near $X_0$ to
'co-moving' fluctuations, simply following the classical trajectory
starting at $X_0$. Save for the replacement $X_0\to X_t$ we have the
same Fokker-Planck equation at all times $t$, except that the
$2\times2$ diffusion matrix becomes time dependent along the classical
trajectory. The Lyapounov rate, on the other hand, neither changes
along the trajectory nor when the point $X_0$ is varied to select
other (infinite) trajectories \cite{Haake}.

The variance ${\rm var}_t(s)=(\overline{s^2})_t-(\overline{s})_t^2$ of
the stable deflection is readily obtained as
\begin{equation}
\label{stablevariance}
{\rm var}_t(s) ={\rm e}^{-2\lambda t}{\rm var}_0(s)+\int_0^t dt'
  {\rm e}^{-2\lambda ( t-t')}\,2D_{ss}(t');
\end{equation}
it must be positive at all times due to the guaranteed existence and
positivity of the $Q$-function. In fact, to make sure we don't fall
victims to our love of poetry, we have numerically checked that the
diagonal element $D_{ss}$ remains mostly positive for a large number
of trajectories of varying energy and coupling parameter. An exemplary
plot of $D_{ss}$ evaluated along one of these trajectories is shown in
Fig.~\ref{fig:Overlap}. The plot exemplifies how $D_{ss}$ only rarely
turns negative. The temporal convolution with
$\e^{-2\lambda t}$ in (\ref{stablevariance}) always entails a positive
variance; see Fig.~\ref{fig:Overlap}.

Most remarkably, the local quantum fluctuations, manifest in the
directions and strengths of diffusion and antidiffusion, 'sniff out'
the asymptotically stable direction $e_s$ of the classical flow,
tuning temselves to make for a positive variance ${\rm var}_t(s) $
and thus a lower quantum bound $\sim \frac 1 {\sqrt j}$ for the scales
accessible to the stable coordinate $s$. Co-motion is crucial since it
lets the quantum fluctuations probe a time span at least
of the order of the Lyapounov time, over which stability
properties of the classical flow become manifest. Reassuringly, the
linearization used to capture the co-moving fluctuations also remains
reliable over that time span.  

 In the unstable
direction, on the other hand, the variance ${\rm var}_t(u)$ keeps
growing indefinitely, as is similarly implied by the reduced
Fokker-Planck equation (\ref{reducedFP}). This is how the $Q$-function gets
smoothed in the classically stable direction while forever extending
its support in the unstable direction. The notorious singular
structures of classical chaos  are thus avoided, and equilibration to the
microcanonical distribution takes place. %\texttt{do we need the
%  footnote that stood here?}

 % \footnote{In a first letter-formate
 %  publication of our work [AAA and FdS] we had illustrated the
 %  interplay of classical and quantum shrinking and stretching on a toy
 %  model with the Fokker-Planck equation (\ref{reducedFP}), albeit with
 %  time independent diffusion coeficient.}

\section{Giant fluctuations due to chaos}
\label{sec:macfluct}
Once more focusing on a range of energy and coupling strength with  predominant
classical chaos, we now turn to revealing giant fluctuations of the
photon number in the stationary microcanical regime. 

If canonical pairs of variables are employed as arguments our
equilibrated $Q$ has the microcanonical form
\begin{equation}
  \label{microcan}
  Q=\frac{1}{\Omega}\delta(h-\epsilon\omega_0) 
\end{equation}
with $\Omega$ the volume of the energy shell and $\epsilon$ the
(dimensionless) energy set by the initial state. That microcanonical
equilibrium will of course be reached from any smooth initial state
with an energy uncertainty similarly negligible as for a coherent
state (where $\frac{\Delta \epsilon}{\epsilon}\sim
\frac{1}{\sqrt{j}}$).  Writing the delta function in the above
distribution amounts to discarding corrections of relative order
$\frac{1}{\sqrt{j}}$.
Stationary means of powers of the
photon number $\langle (a^\dagger a)^m\rangle$ for any integer $m$
are now accessible as moments of the microcanonical density
(\ref{microcan}). In our semiclassical situation, these means are
independent of the ordering of the $2m$ factors $a,a^\dagger$, to
leading order in $j$.  The normally ordered form $\langle
(a^{\dagger m} a^m\rangle$, measurable by absorbtion of the cavity output,
is thus not different from the antinormally ordered form given by the
moments of $Q$ according to (\ref{moments}) nor from the mean powers of
the photon number, $\langle(a^{\dagger } a)^m\rangle$, such that we have
\begin{align}
  \label{microcanmoments}
  \langle (a^\dagger a)^m\rangle=
           j^{m}\int\limits_{I_{\rm min}}^{I_{\rm max}}\!\!dI 
 %\int\limits_0^{2\pi}d\psi\int\limits_{-1}^1d\cos\theta\int\limits_0^{2\pi}d\phi\,
   \!\!\int \!\!d\psi\, d\!\cos\theta\, d\phi\,
                    I^m\,\frac{\delta(h-\epsilon\omega_0)}{\Omega}\,.
\end{align}
The fourfold integral is most easily done in the case of high
energies, $\epsilon>1$, where the whole Bloch sphere is accessible
(cf.~Figs.~\ref{fig:poincarelow},\ref{fig:poincarehigh}). As detailed
in App.~\ref{microaverage}, mean and variance of the photon number
then come out as
\begin{eqnarray}
  \label{eq:2}
  \langle a^\dagger a\rangle&=&\Big(\frac{j}{\omega}\Big)
             \Big(\epsilon\omega_0+\frac{4 g^2}{3\omega }\Big)\\
             \nonumber
  %{\rm Var}(a^\dagger a)=
\langle (a^\dagger a)^2\rangle-\langle a^\dagger a\rangle^2&=&
             \frac{1}{3}\Big(\frac{j}{\omega}\Big)^2
             \Big(\frac{4\epsilon\omega_0 g^2}
               {\omega}+\frac{136g^4}{15\omega^2}+\omega_0^2\Big).
\end{eqnarray}
With the variance of order $j^2$ we indeed confront macroscopic
fluctuations. Notice that to leading order, and up to numerical
factors, the estimate \eqref{boundsasymp} is confirmed. 

We note without presenting calculations that the variance of order
$j^2$ persists down to smaller energies, provided the energy shell is
dominated by chaos.  As visible in Fig.~2, $\epsilon\approx 0.2$
suffices, together with $\gamma\approx 1.5$.

\section{Beyond the Dicke model}
\label{sec:beyond}

Evolution equations with derivatives terminating at second order are
not restricted to the Dicke model. Whenever chaos is generated by a
Hamiltonian of the form of a second-order polynomial in the pertinent
observables and a coherent-state-based $Q$-function can be used, we
expect a Fokker-Planck equation for $Q$. Examples are (i) $SU(3)$
dynamics like the Lipkin model \cite{Gnutzmann:2000uq}, (ii) genuine
many-body systems among which the Bose-Hubbard model
\cite{Trotzky:kx,Roux:2009kx} is of much current interest (here
the Hamiltonian is quartic in annihilation and creation operators, but
due to the absence of antiresonant terms only first and second
derivatives appear in the evolution equation for $Q$), and (iii) kicked systems like
the top \cite{Haake} whose near classical quantum behavior has
recently been observed
experimentally~\cite{Chaudhury:2009fk} (see Appendix~\ref{apptop}).

Even though numerous dynamical systems have Fokker-Planck equations
representing their unitary quantum evolution, this behavior is by no
means generic. In general, the $Q$-function evolves with derivatives
beyond the second order. For non-polynomial Hamiltonians even
infinite-order derivatives appear. The question then arises whether,
given classical chaos, other equilibration mechanisms reign or whether
derivatives with orders $n>2$ give but unimportant corrections to the
quantum diffusion carried by $n=2$. Audacious as general statements
may be we dare pointing to a power counting argument which suggests
prevalence of the mechanism discussed in this paper. When canonical
pairs of variables are used, the generator of the time evolution of
$Q$ has the orders of derivatives and of Planck's constant
interrelated as $\sum_{n=1,2,\ldots}\hbar^{n-1}\partial_X^{n}f_n(X)$
with $X$ and the coefficients $f_n(X)$ independent of $\hbar$. Herein
$n=1$ captures the classical Hamiltonian drift while $n=2$ accounts
for quantum diffusion and brings about the minimal scale $\sqrt \hbar$
for the stable coordinates $s$. We may then set $X\to X_t$ in the
coefficients $f_n$ for $n\geq 2$, integrate out all but the stable
variables, and refer the stable variables to the said quantum scale as
$s=\sqrt\hbar \tilde s$.  A reduced generator appears as
$\partial_{\tilde s} \lambda \tilde s
+\sum_{n=2,,\ldots}\partial_{\tilde s}^{n}\hbar^{(n-2)/2}f_n(X_t) $
and indeed suggests that quantum effects are dominated by the
second-order derivative terms.

Finally, inasmuch as a homogeneously filled energy shell has
macroscopic extent in at least one phase space 'direction',
observables exploring that direction will display macroscopic
stationary fluctuations.

\section{Summary and discussion}
\label{sec:discussion}

The smoothing effect of diffusion on chaotic dynamics has been noted
before, e.g., within the context of quantum billiards
(cf. Ref.~\cite{AlLa1}). However, diffusive contributions to
classical evolution were there added by hand. Our present analysis
exemplifies how unitary quantum evolution itself brings about
diffusion. By projecting the Dicke model dynamics onto a co-moving
Poincar\'e surface of section we could check explicitly that quantum
diffusion sets a limiting scale to the variance of the stable
coordinate such that  $Q$ equilibrates to a smooth
density of the microcanonical form.

It is well to realize that we are facing a privilege of the
$Q$-function which other popular quasi-probability densities like the
Wigner function or the Glauber-Sudarshan $P$-function (weight in a
diagonal mixture of coherent states) do not enjoy. The Wigner function
$W$, for instance, is known to develop positive/negative substructures
within Planck cells under conditions of classical
chaos~\cite{Zurek:2001uq}. Such substructures forbid pointwise
convergence of $W$ to a classical probability density as $\hbar\to 0$;
they are washed out by the average over, roughly, a Planck cell which
leads from $W$ to $Q$.  The situation is even more precarious for the
$P$-function from which $W$ arises by smoothing over, roughly, a
Planck cell. Not only is $P$ prone to going negative but even to
loosing existence as an ordinary function under dynamics with
classical chaos.  For instance, a coherent initial state will get its
support distorted to that of a 'Schr\"odinger cat state' (in the
classicaly unstable direction) \cite {Zurek:2001uq}, which latter is
known to have a non-positive and even singular $P$
\cite{Huang:1996uq}. It is in fact easy to check that for the Dicke
model the diffusion terms $\cal L _{\rm diff}$ for $P$ and $Q$ differ
only in sign (see Appendix~\ref{sec:deriv-quant-evol}); therefore, the
variance ${\rm var}_t(s)$ which remains positive at all times for $Q$
must sooner or later go negative for $P$.

In response to the recent experimental observation of the
superradiant phase transition in the Dicke model, we have investigated
the prospects of detecting the concomitant transition from regular
dynamics at the lowest of energies to prevalence of chaos at higher
excitations. As a most interesting witness of that transition we have
identified  stationary fluctuations of the number of oscillator quanta
(photons). While small for regular dynamics, these fluctuations rise
to macroscopic magnitude as chaos proceeds towards fully covering the
energy shell. Perhaps fortunately for attempts at detection, the
large-fluctuation regime signalling fully chaotic behavior is found
already for moderate degrees of excitation, provided the coupling is
chosen above the critical value for the superradiant phase transition.
The giant fluctuations are predicted to arise independent of the
initial state, after a time of the order of the Ehrenfest time. 

We have also argued that both our equilibration mechanism and large
fluctuations of suitable observables are at work in other observable
systems of current interest.

\textit{Acknowledgments ---} Discussions with T. Brandes, P. Braun,
T. Esslinger, C. Emary, V. Gurarie, M. Ku\'s, J. Larson, and
M. Lewenstein are gratefully acknowledged. Work supported by the
SFB/TR 12 of the Deutsche Forschungsgemeinschaft.

\appendix 

\section{Derivation of the quantum evolution equation}
\label{sec:deriv-quant-evol}
We start the derivation of Eq. \eqref{quantumevol} with the
identity
\begin{align}
\label{eq:6}
   \dot{Q}(\alpha,z) &= {2j+1\over \pi^2(1+|z|^2)^2} \langle \alpha,z| d_t \hat \rho
  |\alpha,z\rangle \cr &=- {\I\over \hbar}  {2j+1\over
    \pi^2(1+|z|^2)^2}\langle\alpha,z| 
    \big[\hat H,\hat \rho\big]
  |\alpha,z\rangle \cr
&= - {\I\over \hbar}  {2j+1\over \pi^2(1+|z|^2)^2}\, {\rm tr}\, \hat\rho
\big[\hat H,  |\alpha,z\rangle \langle\alpha,z|\big] \,.
\end{align}
To process this expression, we need to compute the action of the
Hilbert operators on coherent states. As a result of a straightforward
calculation, one obtains
\begin{align}
\label{eq:11}
  \hat J_-|z\rangle  \langle z| &=\left(\partial_z + 2j{ z^*\over
      1+|z|^2}\right)|z\rangle \langle z|, \cr
 \hat J_+|z\rangle \langle z| &=\left(-z^2\partial_z +2j {  z\over
      1+|z|^2}\right)|z\rangle \langle z|,\cr
 \hat J_z|z\rangle \langle z |&=\left(-z\partial_z + j{ 1-| z|^2\over
      1+|z|^2}\right)|z\rangle \langle z|,
\end{align}
and
\begin{align}\label{eq:12}
  a|\alpha\rangle \langle\alpha| &=\alpha |\alpha\rangle \langle\alpha|,\cr
a^\dagger |\alpha\rangle \langle\alpha| &= \left(\partial_\alpha + 
  \alpha^*\right) |\alpha\rangle \langle\alpha|.
\end{align}
Substituting Eq. \eqref{H} into \eqref{eq:6} and using the relations
above we obtain our Fokker-Planck equation \eqref{quantumevol}.

Likewise, one checks that the Glauber-Sudarshan $P$-function, defined
as the weight in the diagonal mixture of coherent states
\begin{equation}
  \label{Pfunction}
\hat\rho=\int d\alpha dz P(\alpha,z)  |\alpha,z\rangle \langle\alpha,z|\,,
\end{equation}
obeys a Fokker-Planck equation whose generator differs from the one
for the $Q$-function (cf~(\ref{quantumevol})) in only two details: (i) The
factor $(j+1)$ in the drift term in (\ref{quantumevol}) is replaced by
$j$. (ii) Much more importantly, the diffusion term acquires an
overall minus sign.

\section{Transformation to canonical variables}
\label{sec:trafo}

For completeness, we here provide a few technical details relating to
the change of variables $(\alpha,\alpha^\ast,z,z^\ast)\to
(I,\psi,c\equiv \cos\theta,\phi)$. The defining relations
\begin{equation}
  \label{eq:4}
  z= \tan\frac{\theta}{2} \,\e^{\I\phi}=\sqrt{\frac{1-c}{1+c}} 
  \e^{\I\phi} \qquad 
  \alpha=\sqrt{jI}\e^{\I\psi}
\end{equation}
yield the derivatives
\begin{eqnarray}
  \label{partials}\nonumber
  \partial_z&=&-\frac{1}{2}\left[\partial_c (1+c)\sqrt{1-c^2}
                  +\I\partial_\phi\sqrt{\frac{1+c}{1-c}}\right]\e^{-\I\phi}\\
                  &&+\sqrt{1-c^2} \e^{-\I\phi} \,,
  \\ \nonumber
  \partial_\alpha&=&\Big[\partial_I\sqrt{\frac{I}{j}}-\frac{\I}{2\sqrt{jI}}\Big] \e^{-\I\psi}
\end{eqnarray}
and their complex conjugates. We must realize that the complex
stereographic projection variables $(z,z^*)$ are
not a canonical pair. Therefore, a
Jacobian arises in $\tilde Q(c,\phi,I,\psi)=\frac{2}{(1+c)^2}Q(z(c,\phi),
z^*(c,\phi),\alpha(I,\psi),\alpha^*(I,\psi))$ such that to get the
generator for $\tilde Q(c,\phi,I,\psi)$ we must replace as $\partial_c
\to (1+c)^{-2}\partial_c (1+c)^{2}=\partial_c+2 (1+c)^{-1}$ and
\begin{equation}
  \label{eq:13}
  \partial_z \to -\frac{1}{2}\Big[\partial_c (1+c)\sqrt{1-c^2}
                  +\I\partial_\psi\sqrt{\frac{1+c}{1-c}}\Big]\e^{-\I\phi}\,.
\end{equation}
Straightforward calculation then gives the generator in search as
$\partial_{X_i}d_i+\partial_{X_i}\partial_{X_j}D_{ij}$ with the drift vector
\begin{eqnarray} \label{driftvector}
     d_I&=&\sqrt{8}
     g\Big(1+\frac{1}{j}\Big)\sqrt{I} \sin\psi\sqrt{1-c^2} \cos\phi
   \\  \nonumber
    d_\psi& =& \left[\omega+\sqrt{2}g\Big(1+\frac{1}{j}\Big)\frac{1}{\sqrt{I}}
                          \cos\psi \sqrt{1-c^2}\cos\phi\right]
    \\  \nonumber
    d_c&=& \left[-\omega_0+\sqrt{8}g\sqrt{I}\cos\psi\frac{c}{\sqrt{1-c^2}}
                                  \cos\phi\right]
   \\ \nonumber
    d_\phi&=& -\sqrt{8}g\sqrt{I}\cos\psi  \sqrt{1-c^2}\sin\phi
    \,,\\ \nonumber          
\end{eqnarray}
and the diffusion matrix
\begin{eqnarray}
  \label{diffmatrix}
  D&=&\left({0\atop d^\dagger}{d \atop 0}\right)\,,\\ \nonumber
  d&=&\left({D_{Ic}\atop D_{\psi c}}{D_{I\phi} \atop
      D_{\psi\phi}}\right) \\ \nonumber
    &=&\frac{g}{j\sqrt{2}}\left({\sqrt{I}\atop 0}{0 \atop \frac{1}{2\sqrt{I}}}\right)
           \left({A \atop -B}{B \atop A}\right)
           \left({\sqrt{1-c^2} \atop 0}{0 \atop
               \frac{1}{1-c^2}}\right)\,,\\ \nonumber
    A&=&-\cos\psi\sin\phi-c\,\sin\psi \cos\phi\,,\\ \nonumber
    B&=&c\,\cos\psi \cos\phi-\sin\psi\sin\phi \,.
\end{eqnarray}

\section{Bounds on photon number in energy shell}
\label{bounds}

Solving Eq.~\eqref{eq:10} for $I$, one readily finds
\begin{align}
  \label{eq:9}
  \frac{\sqrt{2\nu I}}{\gamma} = %{\gamma\over \sqrt{2\nu}}
  \pm\sqrt{(x\sin\theta)^2+{2\over
        \gamma^2}\left(\epsilon-\cos\theta\right)}-x\sin\theta\,,
\end{align}
where we defined $x=\cos\psi \cos\phi$. Interestingly, as a function
of $\epsilon, \cos\theta, \phi, \psi$, the quantity $I$ may take on
two different values, and that complication arises when
both $x$ and $\epsilon-\cos\theta$ are negative. No solution for $I$
exists when non-negative $x$ meets with negative
$\epsilon-\cos\theta$.  The simplest situation is
$\epsilon-\cos\theta>0$: then only the positive square root in
(\ref{eq:9}) is possible and $I$ is unique. We illustrate the search
for the bounds with just a few cursory remarks on the latter case.

There is no extremum in the calculus sense. So the smallest and
largest values of $I$ must occur on the boundaries $\cos\theta=\pm1$
or/and $x=\pm1$. The poles of the Bloch sphere provide the "trivial" bounds
$I_{\rm min \atop \rm max}=\nu^{-1}(\epsilon \pm 1)$. 

To check the possibility of tighter bounds we first try $x=1$. From
$\partial \sqrt{I}/\partial\cos\theta=0$ we get
\begin{align}
  \label{bounds1}
  \cos\theta=\frac{\gamma^2\pm\sqrt{\gamma^4+1+2\gamma^2\epsilon}}
                        {1+2\gamma^2\epsilon}\,, \;%\cr
  % \sin\theta&=&\frac{\sqrt{ (1+2\gamma^2\epsilon)^2
  %                        -(\gamma^2\pm\sqrt{\gamma^4+1+2\gamma^2\epsilon})^2}}
  %                       {1+2\gamma^2\epsilon}\\ \label{bounds3}
   \sqrt{I}={\gamma\over \sqrt{2\nu}}\tan\theta\,.
\end{align}
Due to $\sqrt{I}\geq 0$ and the global positivity of $\sin\theta$, we
must require $\cos\theta\geq 0$ and therefore only the upper sign
qualifies. A non-trivial lower bound $I_{\rm min}$ is thus obtained. Similarly,
the case $x=-1$ yields a non-trivial upper bound  $\sqrt{I_{\rm
    max}}=-{\gamma\over \sqrt{2\nu}}\tan\theta$ with $\tan\theta$
according to the lower sign in (\ref{bounds1}).  The leading terms of
the $\epsilon$-expansion of these bounds are the ones given in
(\ref{boundsasymp}).

\section{Microcanonical averages}
\label{microaverage}

We briefly sketch the calculation of the microcanonical moments
(\ref{microcanmoments}), for simplicity confining ourselves to high
energies ($\epsilon>1$).  Doing the $I$-integral in
(\ref{microcanmoments}) we have
\begin{align*}
  M_m\equiv\langle (aa^\dagger)^m\rangle=
           j^{m}\frac{1}{\Omega}\int\!d\psi\, d\!\cos\theta\, d\phi\,
             \,\frac{\hat{I}^m}{\left|\partial h/\partial
                 I\right|_{I=\hat I}},
\end{align*}
with the peak intensity $I=I(\psi,\phi,\theta)$ determined by
\eqref{eq:9} and $x=\cos\psi \cos\phi$; only the positive sign in
\eqref{eq:9} is possible at high energies.  The derivative of the
Hamiltonian
\begin{align*}
  \frac{1}{\left|\partial h/\partial I\right|}=
    \frac{ 2}{\omega_0\sqrt{2\nu} \gamma}
     \frac {I}
        {\sqrt{(x\sin\theta)^2+{2\over \gamma^2}(\epsilon -\cos\theta)}}\,.   
\end{align*}
allows to rewrite  %the above expression for 
the moments as
\begin{align*}
 M_m&=\Big(\frac{j\gamma^2}{2\nu}\Big)^m\frac{1}{\nu\omega_0\Omega}
           \int\!d\psi \,d\!\cos\theta \,d\phi \cr
&\frac{\left(
   \sqrt{(x\sin\theta)^2+{2\over \gamma^2}(\epsilon-\cos\theta)}-x\sin\theta\,\right)^{2m+1}}
        {\sqrt{(x\sin\theta)^2+{2\over \gamma^2}(\epsilon-\cos\theta)}}\,.
\end{align*}
The power in the foregoing numerator can be binomially expanded. By
symmetry only even powers of $x\sin\theta$
contribute and therefore only even powers of the square root
remain. We quickly find 
\iffalse
$M_0=1=\frac{2(2\pi)^2}{\nu\omega_0\Omega}$ and thus
\begin{align*}
 M_m&=\frac{1}{2(2\pi)^2}\Big(\frac{j\gamma^2}{2\nu}\Big)^m
               \sum_{n=0}^m\left(\!{2m+1\atop 2n}\!\right)I_{mn},\cr
   I_{mn}&= \int\!d\psi \,d\!\cos\theta \,d\phi\,(x\sin\theta)^{2n}\cr
       &\times
 \left((x\sin\theta)^2+{2\over \gamma^2}(\epsilon-\cos\theta)\right)^{m-n}\,.
\end{align*}
The remaining integrals are elementary and yield
\fi
\begin{align*}
 M_1&=\Big(\frac{j}{\nu}\Big)\,\big(\epsilon+\textstyle{\frac{1}{3}}\gamma^2\big)
               \cr
  M_2&=\Big(\frac{j}{\nu}\Big)^2\,\big(\epsilon^2+\epsilon\gamma^2
                   +\textstyle{\frac{3}{10}}\gamma^4+\textstyle{\frac{1}{3}}\big)\,,
\end{align*}
which immediately implies \eqref{eq:2}.

\section{Kicked top}
\label{apptop}
We would like to corroborate our expectation for the \textit{kicked
  top} \cite{Gnutzmann:2000uq,Haake:2010fk}, a periodically kicked large spin with conserved length,
$\vec{\hat{J}}^2=j(j+1)\gg 1$. Classical equilibration for a cloud of
points on the Bloch sphere has long been known from numerical
studies. Quantum equilibration and the ensuing large stationary
fluctuations of the (orientation of the) angular momntum should be
observable in a variant of the experiment of Ref.~\cite{Chaudhury:2009fk}.

The simplest chaotic top has the Floquet operator 
\begin{equation}\nonumber
  %\label{topFloquet}
  \hat F=\e^{-\I\frac{\tau}{2j+1} \hat J_z^2}\,\e^{-\I p\hat J_x}\,;
\end{equation}
it involves a rotation about the $\hat J_x$-axis by the angle $p$
and a subsequent 'torsion' about the $\hat J_z$-axis. Torsion means a
state dependent rotation by the angle $\frac{\tau}{2j+1} \hat J_z$
which has opposite signs in the northern and
southern hemispheres. The precession angle $p$ and the torsion
constant $\tau$ are assumed independent of $j$. Chaos predominates if
$\tau\gg 1$.

The stroboscopic time evolution of the density operator is given by
$\hat\rho_n=\hat F^n\hat\rho_0 \hat F^{\dagger n}$ with the dimensionless
integer 'time' $n$. Employing the $Q$-function $Q(z)=\frac{2j+1}{\pi
  (1+zz^*)^2}\langle z|\hat\rho|z\rangle$ we go for the propagator for
its single-step evolution $Q_{n+1}(z)=\mathcal F Q_n(z)$. Like the
Floquet operator $\hat F$, the Husimi propagator $\mathcal
F=\mathcal F_\tau \mathcal F_p$ is a product of two factors, one each
for precession and torsion.

For the precession we get
\begin{equation}\nonumber
  %\label{Fp}
  \mathcal F_p=\exp\left\{\textstyle{\frac{1}{2}}
                       \I p\partial_z (1-z^2) + \mathrm{c.c.}\right\}\,.
\end{equation}
The generator in the foregoing exponent involves only drift (first
order derivative terms) but no diffusion; it is the generator
for rotation about the $J_x$-axis already encountered for the Dicke
model in (\ref{quantumevol}), classical Hamiltonian in character. The
torsion propagator reads
\begin{equation}\nonumber
  %\label{Ftau}
  \mathcal F_\tau=\exp\!
 \left\{\!-\I\tau\!\left[\partial_zz\frac{1-zz^*}{1+zz^*}+\ldots \right]
         +\frac{\I\tau}{2j+1}\partial_z^2z^2+\mathrm{c.c.}\!\right\}\!,
\end{equation}
with an exponent involving drift and diffusion. A
$\frac{1}{j}$-correction in the drift has not been written out. The displayed
drift is again classical Hamiltonian, as becomes visible once the real
canonical pair $\cos\theta, \phi$ of variables is introduced. The
quantum diffusion with an explicit factor of order $\frac{1}{j}$
involves a real symmetric $2\times 2$ diffusion matrix with
vanishing diagonal elements (i.~e., of chiral structure).

All arguments for equilibration for the Dicke model apply again.  Most
importantly, the positive eigenvalue of the diffusion matrix sets a
smallest scale for the motion along the stable direction of the
classical drift. On that latter scale $Q$ becomes smooth transverse to
the classically unstable direction. Effective stationarity will reign
no later than about an Ehrenfest time, with $Q$ constant over the Bloch
sphere. Equipartition of $Q$  will result in large
stationary fluctuations of the angular momentum.

%\bibliography{my}
\bibliography{/Users/alex/Documents/REVTEX/BIBLIOGRAPHY/my}
\bibliographystyle{apsrev4-1}

% \begin{thebibliography}{99}

%\bibitem{Glauber} R. Glauber, {\it Quantum Theory of Optical
%Coherence}, Selected Papers and Lectures, Wiley, Weinheim (2007) 

%\bibitem{Arecchi} F.T. Arecchi, E. Courtens, R. Gilmore, and
%H. Thomas, Phys. Rev. A 6, 2211 (1972)

%\bibitem{RGFH} R. Glauber and F. Haake, Phys. Rev. A 13, 357 (1973) 

% \bibitem{Larson} J. Larson and M. Horsdal, Phys. Rev. A 84,
% 021804(R) (2011)

%\bibitem{Goldstein} S. Goldstein, J.L. Lebowitz, C. Mastrodonato, R. Tumulka, and
%N. Zanghi, Phys. Rev E. 81, 011109 (2010)

%\bibitem{Zurek} W. H. Zurek, Nature 412 (2001)

%\bitem{Huang} H. Huang, S.-Y. Zhu, M.S. Zubairy, Phys. Rev. A 53,
%1027 (1996)

%\bibitem{HaakeKus} F. Haake, M. Ku\'s, Scholarpedia
\end{document}